\documentclass{PoS}

\usepackage{graphicx}

\title{RHIC Critical Point Search: Assessing STAR's Capabilities}

\ShortTitle{STAR's Capabilities}

\author{Paul Sorensen for the STAR Collaboration\\
        Brookhaven National Laboratory\\
        E-mail: \email{prsorensen@bnl.gov}}

\abstract{ In this report we discuss the capabilities and limitations
  of the STAR detector to search for signatures of the QCD critical
  point in a low energy scan at RHIC. We find that a RHIC low energy
  scan will cover a broad region of interest in the nuclear matter
  phase diagram and that the STAR detector --- a detector designed to
  measure the quantities that will be of interest in this search ---
  will provide new observables and improve on previous measurements in
  this energy range. }

\FullConference{The 3rd edition of the International Workshop --- The Critical Point and Onset of Deconfinement --- \\
		 July 3-7 2006\\
		 Galileo Galilei Institute, Florence, Italy}

\begin{document}

Experiments at RHIC have found evidence that a strongly coupled
quark-gluon plasma is created in heavy-ion collisions at
$\sqrt{s_{_{NN}}}=200$~GeV~\cite{whitepapers}. At these high energies
the baryon chemical-potential ($\mu_{B}$) extracted from thermal
model fits is small (approximately 0.025 GeV)~\cite{cleymans}. Lattice
calculations indicate that for $\mu_B=0$, as the temperature ($T$) of
nuclear matter is increased the transition from confined to deconfined
matter is a smooth crossover~\cite{Brown:1990ev} and that the chiral
and deconfinement transitions happen at approximately the same
temperature~\cite{Mocsy:2003qw, Karsch:1998qj}. Model calculations,
however, suggest that for $T=0$, the transition as $\mu_B$ is
increased is first order~\cite{models}. If this is the case, then a
critical point should exist where the transition changes from first
order, to a smooth-crossover~\cite{Stephanov:2004wx}.

\begin{figure}[htb]
\centering\includegraphics[width=0.7\textwidth]{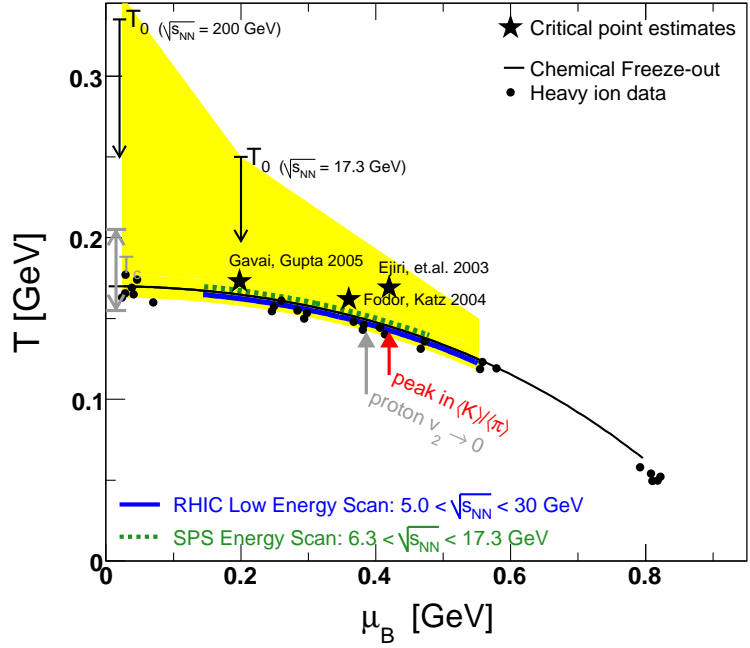}
\caption{ A sketch of the phase diagram of nuclear matter showing the
  chemical freeze-out curve, points from fits to heavy-ion collision
  data, estimates of the initial temperatures achieved in heavy-ion
  collisions, lattice QCD estimates of the critical temperature $T_C$
  for $\mu_B=0$~\cite{lattTC}, and lattice QCD estimates of the
  location of the critical point~\cite{lattCP}. The shaded region
  shows an estimate of the region of the phase diagram that can be
  covered by a low energy scan at RHIC. }
\label{fig1}
\end{figure}

It may be possible to ascertain the ($T$, $\mu_B$) coordinates of the
critical point by decreasing the collision energy for heavy-ion
collisions at RHIC~\cite{rbrcmeeting}: $\mu_B$ increases with
decreasing $\sqrt{s_{_{NN}}}$. A non-monotonic dependence of variables
on $\sqrt{s_{_{NN}}}$ and an increase in event-by-event fluctuations
should become apparent near the critical
point~\cite{Stephanov:1998dy,Ejiri:2005wq}. The energy scan at the
CERN-SPS ($6.3 < \sqrt{s_{_{NN}}} < 17.3$~GeV) found some possible
signatures of a critical point, but the evidence remains inconclusive
and sometimes
contradictory~\cite{Afanasiev:2002mx,Alt:2003ab,Roland:2004pu}. Using
a collider to perform such an energy scan, instead of fixed-targets,
should provide two important advantages: acceptance won't change with
$\sqrt{s_{_{NN}}}$ and track-density at mid-rapidity will only vary
slowly~\cite{Roland:rbrc}. In addition, the detectors at RHIC are of a
more advanced design~\cite{Ludlam:2003sn}. Figure~\ref{fig1} shows
lattice QCD estimates of the critical temperature $T_{C}$ for
$\mu_B=0$~\cite{lattTC}, lattice QCD estimates of the location of the
critical point~\cite{lattCP} and an estimate of the region that can be
covered by a RHIC low energy scan. We find that the RHIC low energy
scan will cover a broad region of interest in the $T$, $\mu_B$
plane~\cite{T0}.

In this report, we will assess the capabilities of the STAR
detector~\cite{STAR} at RHIC to perform a critical point search in
which the center-of-mass energy of collisions may be reduced to as low
as $\sqrt{s_{_{NN}}} \sim 4.5$~GeV. We focus on several key
measurements: yields, fluctuations in particle ratios, and elliptic
flow. The STAR detector was designed for these measurements so we
expect it to perform well. We present simulations that indicate this
is the case. We also indicate where potential difficulties may arise.

\section{STAR Detector}

\begin{figure}[htb]
\includegraphics[width=1.0\textwidth]{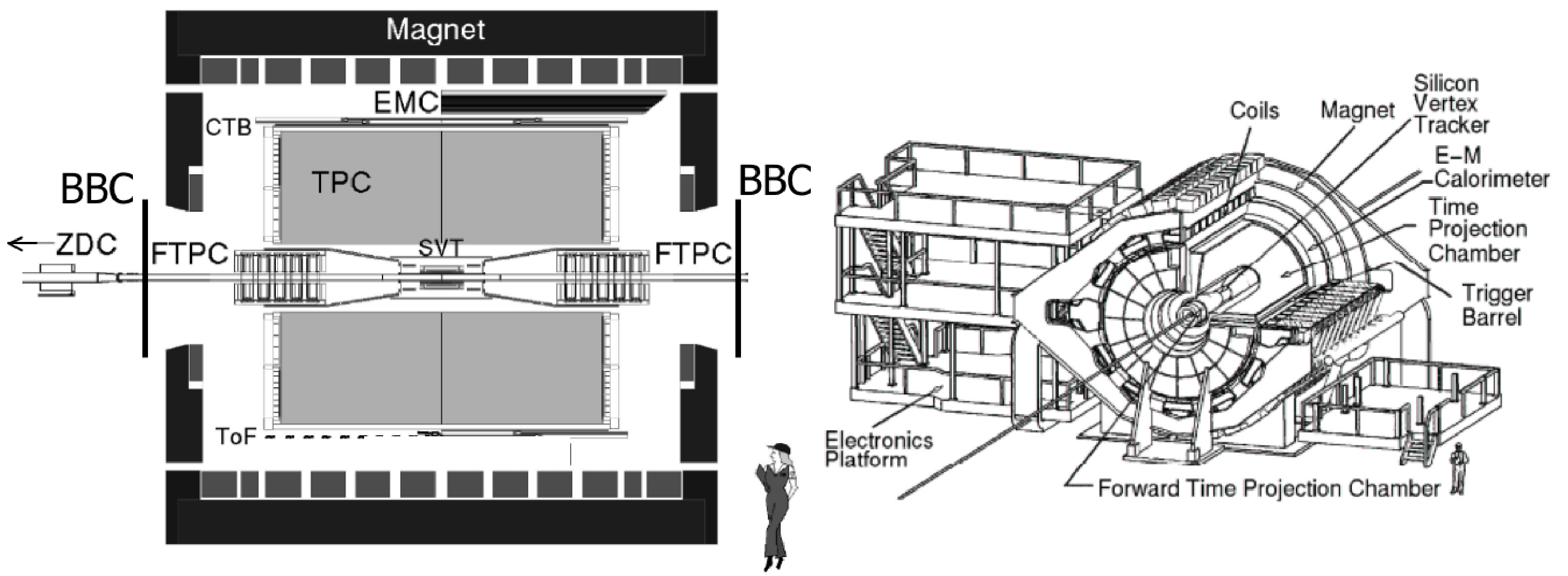}
\caption{The STAR detector~\cite{STAR}: a cross-section view (left
  panel) and perspective view (right panel). STAR an azimuthally
  symmetric, large acceptance, solenoidal detector designed to measure
  many observables simultaneously. The detector consists of several
  subsystems and a large Time Projection Chamber (TPC)~\cite{tpc}
  located in a 0.5 Tesla solenoidal analyzing magnet.}
\label{fig2}
\end{figure}

The layout of the STAR detector system as it was for Run-2 is shown in
Figure~\ref{fig2}. The active subsystems included two RHIC-standard
zero-degree calorimeters (ZDCs) that detect spectator neutrons, a
central trigger barrel (CTB) that measures event multiplicity, a
time-of-flight detector, an electromagnetic calorimeter to measure
photons, electrons and the transverse energy of events, and four
tracking detectors. The tracking detectors are the main TPC, two
forward TPCs, and the silicon vertex tracker (SVT).

The TPC is STAR's primary detector~\cite{tpc} and can track up to
$\sim 4 \times 10^3$ particles per event. For collisions in its
center, the TPC covers the pseudo-rapidity region $|\eta| < 1.8$.  It
can measure particle $p_T$ within the approximate range $0.07 < p_T <
30$~GeV/c. The momentum resolution $\delta p/p$ depends on $\eta$ and
$p_T$ but for most tracks $\delta p/p \sim 0.02$. The full azimuthal
coverage of the STAR detector ($-\pi < \phi < \pi$) makes it ideal for
detecting weak decay vertices, reconstructing resonances, measuring
$v_2$ and event-by-event fluctuations.

\begin{figure}[htb]
  \resizebox{0.52\textwidth}{!}{\includegraphics{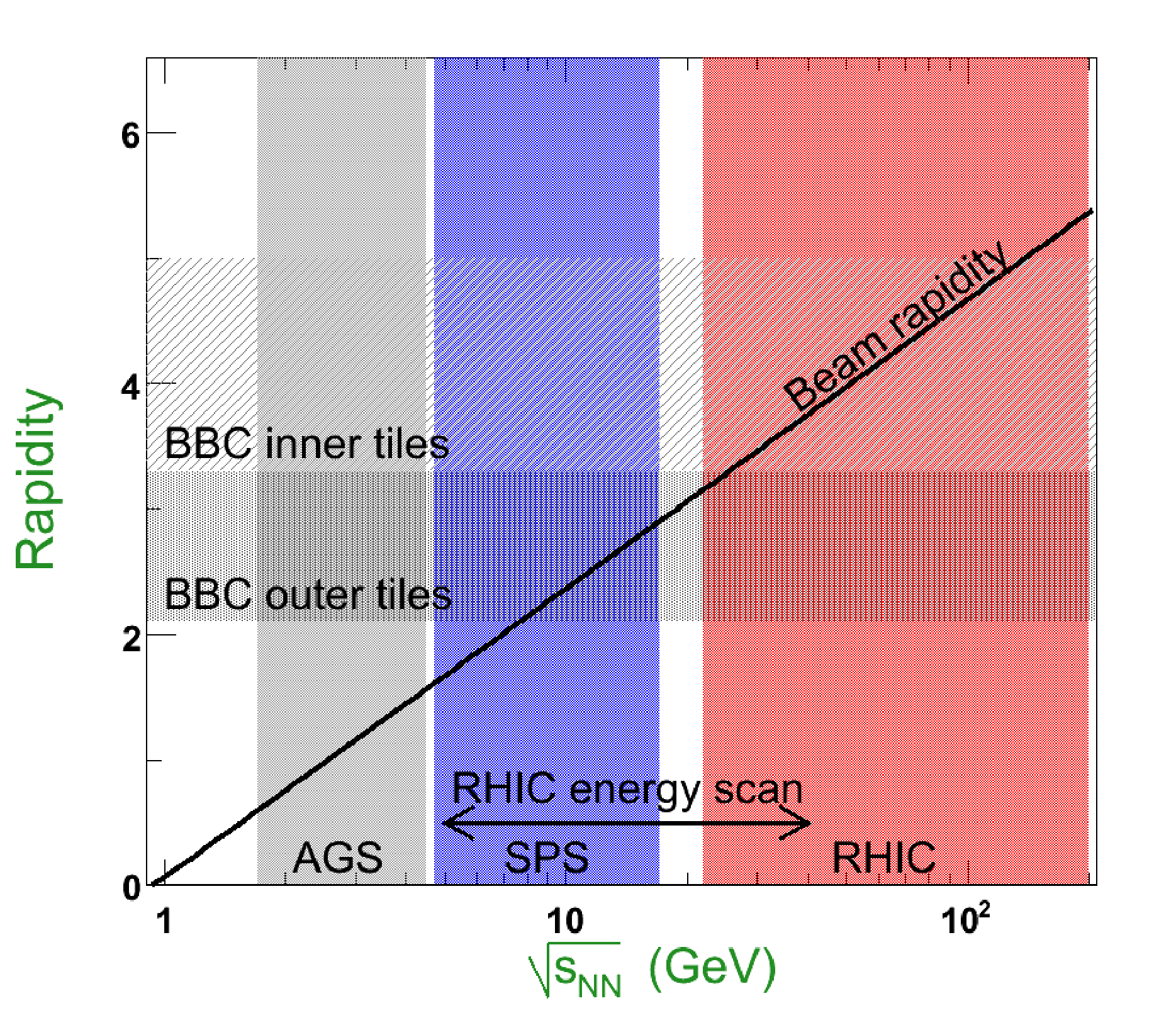}}
  \resizebox{0.48\textwidth}{!}{\includegraphics{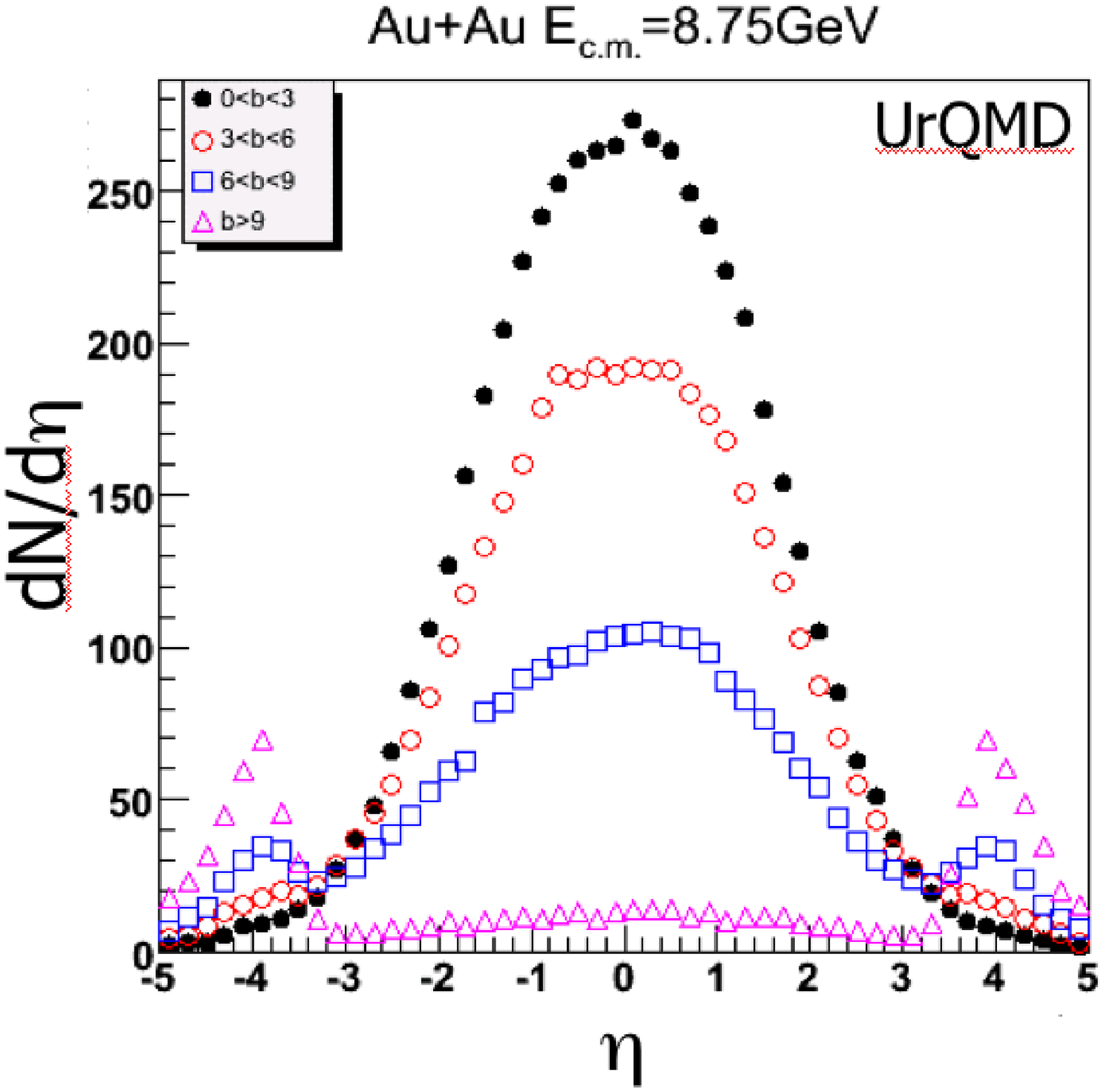}}
  \caption{Beam rapidity versus $\sqrt{s_{_{NN}}}$ (left panel) and
    the corresponding acceptance for STARs Beam-beam counters. UrQMD
    simulation of charged hadron yields for
    $\sqrt{s_{_{NN}}}=8.75$~GeV (right panel).}
\label{fig3}
\end{figure}

The read-out rate for the STAR TPC is $\sim 100$ Hz. For
$\sqrt{s_{_{NN}}}<20$~GeV, without electron cooling upgrades to RHIC,
the bunch crossing and event rates will be much slower than the TPC
read-out rate. As such, STAR will be able to record every detected
event. To trigger on the occurrence of a collision, STAR will need to
use it's Beam-Beam Counters (BBC). In Figure~\ref{fig3} (right)
UrQMD~\cite{UrQMD} simulations of the multiplicity of tracks versus
pseudo-rapidity are shown. Although the Zero-degree Calorimeters will
not register a signal for collisions with $\sqrt{s_{_{NN}}}~<20$~GeV,
tracks will impinge on STARs BBCs. Table~\ref{tab1} lists the number
of charged particles within the BBC acceptance for several energies
and centralities. In all cases, the number of tracks is sufficient to
detect the occurrence of a collision.

\begin{table}[hbt]
\caption{Number of particles impinging on STARs Beam-Beam Counters for
  Au+Au collisions in UrQMD~\cite{UrQMD} simulations.}
\label{tab1}
\centering\begin{tabular}{c|cc|cc}
%\colrule
impact & \multicolumn{2}{c|}{$\sqrt{s_{_{NN}}}=5$~GeV} & \multicolumn{2}{c}{$\sqrt{s_{_{NN}}}=8.75$~GeV} \\
 parameter (fm) & BBC Inner & BBC Outer & BBC Inner & BBC Outer \\
\hline
0<b<3   &   5 & 27  & 12 & 54 \\
3<b<6   &  11 & 30  & 21 & 57 \\
6<b<9   &  22 & 35  & 39 & 40 \\
\hline
\end{tabular}
\end{table}

STARs ability to effectively trigger on events at these low energies
coupled with our proven ability to statistically extract particle
identification information over a broad $p_T$~\cite{Shao:2005iu} range
will make measurements of particle spectra and ratios possible. As
such, $T$ and $\mu_B$ can be extracted from statistical model fits. We
note, however, that many of these particle identification methods are
statistical and therefore cannot be used to identify particles on an
event-by-event basis: as will be needed for particle ratio fluctuation
measurements. For event-by-event particle identification over a broad
$p_T$ range, a Time-of-Flight detector~\cite{TOF} is being constructed
that will cover $2 \pi$ in azimuth and $-1<\eta<1$ in
pseudo-rapidity. This detector upgrade is expected to be finished by
2010.

\section{Analysis}

The STAR detector will be able to improve on many of the measurements
of interest in a critical point search. Since event rates are not
expected to be particularly large (approximately 5 Hz at
$\sqrt{s_{_{NN}}}=5$~GeV), rare probes such as the $J/\psi$ will
likely be inaccessible. Below, we discuss some of the key measurements
for a low energy scan at RHIC. In these proceedings we do not discuss
in detail such important measurements as HBT~\cite{hbt},
$v_1$~\cite{starv1}, balance functions~\cite{balance} and
multiplicity, net charge, or $\langle p_T\rangle$
fluctuations~\cite{starfluct}: all of which are measurements STAR will
be able to perform well. Rather, we present a subset of measurements
--- $v_2$, $v_2$ fluctuations, and dynamical fluctuations of the
kaon-to-pion ratio --- in order to illustrate the STAR detectors
capabilities for a critical point search at RHIC.

\begin{figure}[hbt]
  \resizebox{0.5\textwidth}{!}{\includegraphics{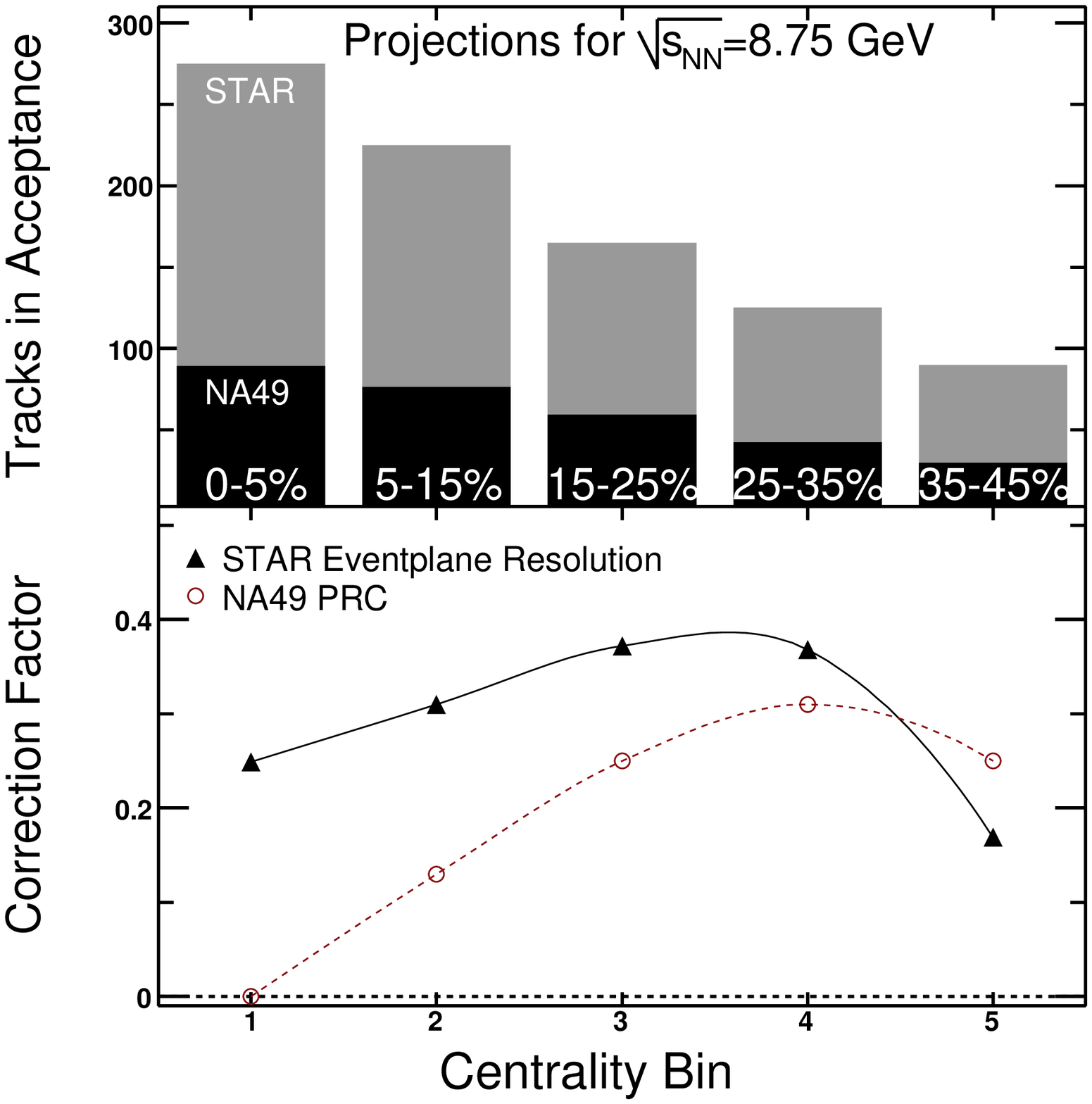}}
  \resizebox{0.5\textwidth}{!}{\includegraphics{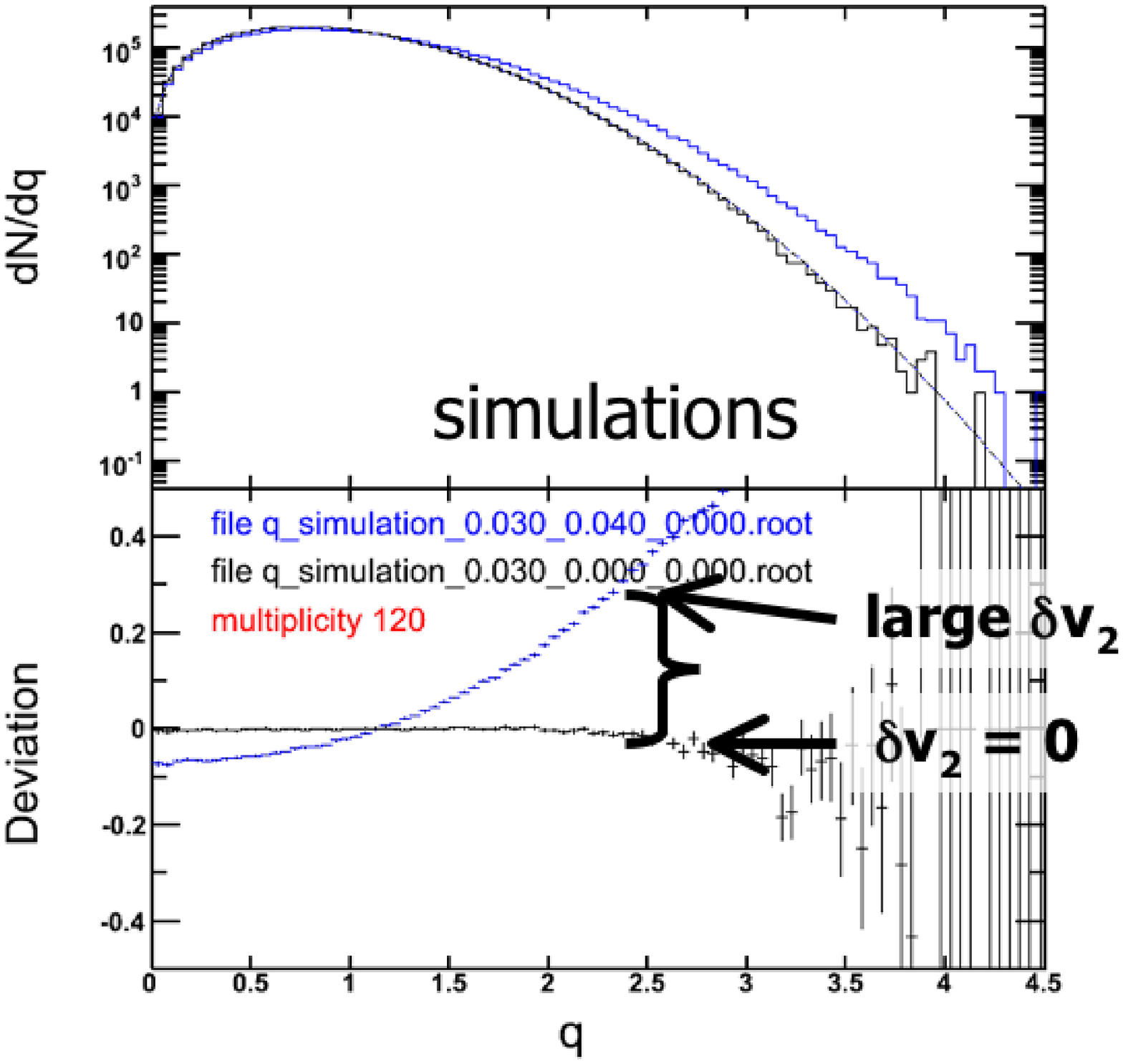}}
  \caption{Top left: The expected number of tracks within the STAR TPC
    acceptance and simulations of the event-plane resolution
    correction factor (bottom left). The left figure also shows the
    corresponding NA49 quantities for
    comparison~\cite{Alt:2003ab}. Top right: The expected shape of the
    distribution of the length of the $q$-vector (curve) for the given
    values of $v_2$ and simulation results (histograms) with and
    without $v_2$ fluctuations. Bottom right: Deviations between
    simulations and the expected shape due to $v_2$ fluctuations
    demonstrating that the distribution of the length of the
    $q$-vector distribution is sensitive to $v_2$ fluctuations. }
\label{fig4}
\end{figure}

A key indicator of the ability to measure elliptic flow is the
reaction-plane resolution correction factor~\cite{v2meth}. When the
factor is large (close to unity), the direction of the reaction plane
can be well determined on an event-by-event basis and fewer events
will be needed for an accurate determination of $v_2$. The resolution
depends on the number of tracks used and the magnitude of the event
asymmetry. For the most peripheral events the small number of tracks
available reduces the resolution while for the most central events the
symmetry of the collision overlap region degrades it. For Au+Au
collisions, the resolution is typically greatest at a centrality
corresponding to roughly 20--30\% of the collision
cross-section~\cite{starv2}. In Fig.~\ref{fig4} (bottom left) we show
the expected resolution correction factor for Au+Au collisions at
$\sqrt{s_{_{NN}}}$ = 8.75 GeV in the STAR detector along with the
resolution correction factor achieved by the NA49 collaboration. The
observed $v_2$ value must be divided by the resolution correction
factor to get the true $v_2$. When this factor is closer to zero than
one, both the observed $v_2$ and the statistical errors must be scaled
up by a large number. A large improvement is expected with the STAR
detector due to it's full azimuthal coverage extending over 2 units of
pseudo-rapidity. This leads to a large increase in the number tracks
available for the measurement of $v_2$. Figure~\ref{fig4} (top left)
shows the number of tracks available for analysis by NA49 and STAR.

In the most peripheral collisions the resolution correction factor for
NA49 is apparently larger than our simulation results for STAR. Given
the larger coverage of the STAR detector, it is unlikely that the NA49
detector can have a better resolution. Either the STAR resolution is
under-estimated for this bin or the NA49 measurements lose accuracy
where the multiplicity is at its lowest value.

The right panel of Fig.~\ref{fig4} shows the distribution of the
length of the flow vector $q$ from simulations~\cite{myqm06}. The
histograms in the top panel are the simulated data with and without
$v_2$ fluctuations.  The curve shows the expected shape of the
distribution derived from the central limit theorem~\cite{qdist}. The
difference between the two histograms shows how much the distribution
changes if $v_2$ fluctuates from event-to-event. The bottom panel
shows deviations between the simulated data and the expected
shape. The simulations use $v_2=0.03$, and a multiplicity of 120
tracks. $v_2$ fluctuations in the simulation are Gaussian with either
zero width or a width of $0.04$. The figure demonstrates that $v_2$
fluctuations can be measured using the distribution of the length of
the flow vector. This measurement will remove a dominant source of
systematic uncertainty in $v_2$ measurements and provide a robust
critical point signature. Preliminary results have already been
presented~\cite{myqm06,Alver:2006pn}.

\begin{figure}[hbt]
\includegraphics[width=1.0\textwidth]{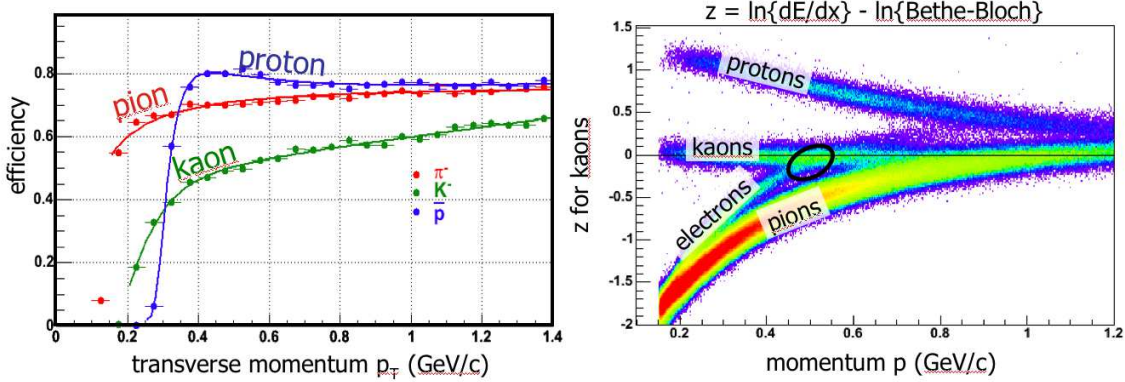}
\caption{Left panel: The efficiency versus $p_T$ for detecting pions,
  kaons, and protons with the STAR detector. Right panel: The
  $z$-variable for kaons derived from ionization energy loss
  measurements $dE/dx$ in the STAR TPC. This variable illustrates how
  well a particular particle species can be distinguished from other
  particles.}
\label{fig5}
\end{figure}

The left panel of Fig.~\ref{fig5} shows the STAR detectors efficiency
for detecting pions, kaons, and protons versus $p_T$. The efficiency
for detecting kaons is lower than for pions or protons. This reduction
is due to the failure to detect kaons that decay before they traverse
the TPC volume. In a fixed target experiment, the momentum boost in
the lab frame makes it much less likely that a weak-decay will occur
before the kaon has traversed the detector. The efficiency near the
kaon $\langle p_T \rangle$ is approximately 45\%.

The right panel of Fig.~\ref{fig5} shows the $z$-variable for kaons
which illustrates the ability of STAR to distinguish between kaons and
other particle species using ionization energy loss measurements
$dE/dx$ in the TPC volume. $z$ is the logarithm of the ratio of the
measured $dE/dx$ to the expected $dE/dx$~\cite{dedx} for a particular
particle species. Kaons can be identified with good certainty when the
particle momentum is below 400 MeV. Above that value though, electrons
begin to contaminate the kaon sample. Many electrons come from pion
decays so they need to be excluded from the kaon sample in order to
accurately extract the kaon-to-pion ratio event-by-event. In addition,
since pions can yield more than one electron in their decay chain,
mixed events may not be able to account for electron contamination of
the kaon sample. Requiring a good purity for the kaon sample will
further reduce the efficiency for kaon detection: typical kaon
efficiency values in STAR kaon-to-pion fluctuation analyses can be as
low as 10\%--15\%~\cite{kpianal}. For this reason, these measurements
are challenging.

\begin{figure}[htb]
  \resizebox{0.50\textwidth}{!}{\includegraphics{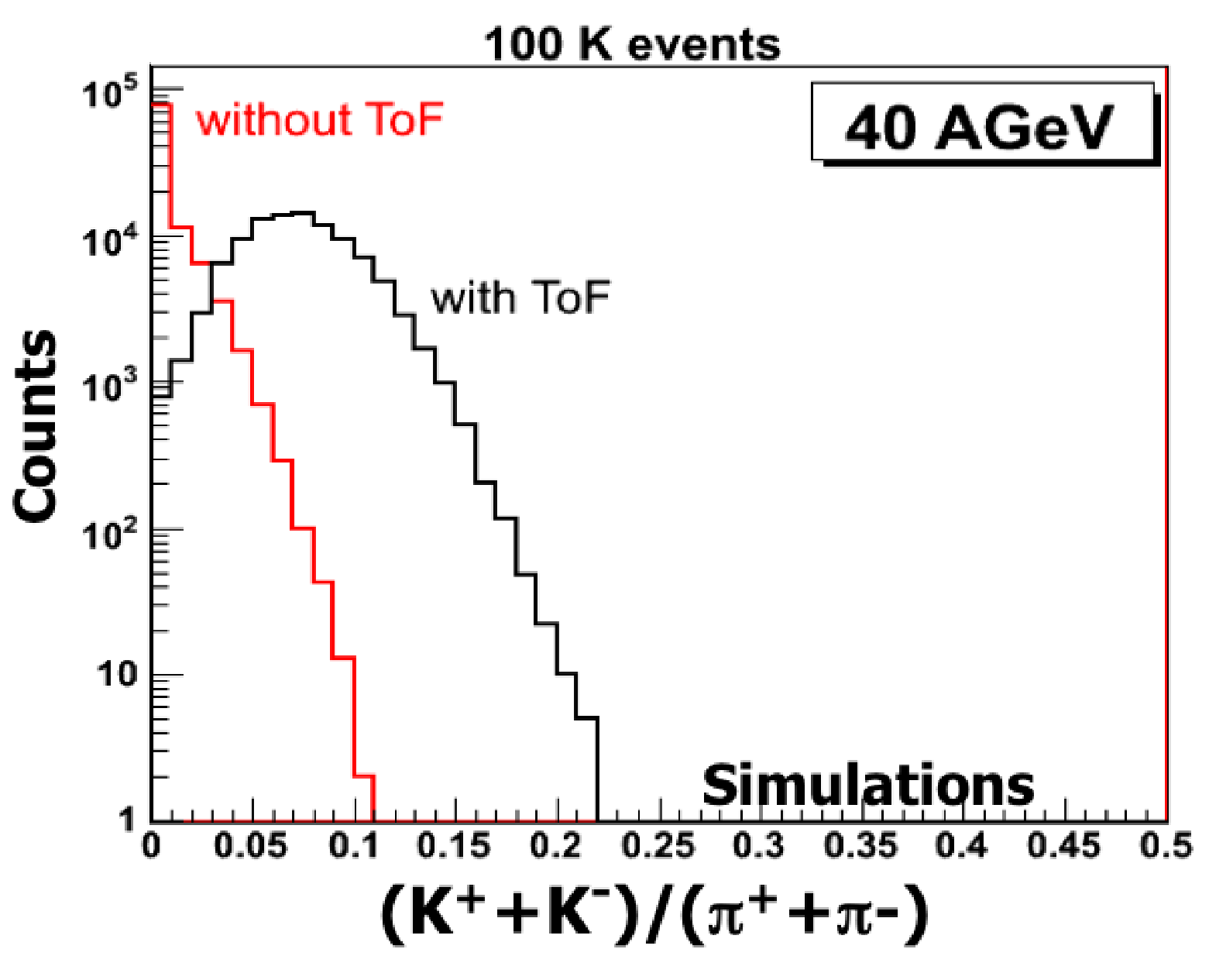}}
  \resizebox{0.50\textwidth}{!}{\includegraphics{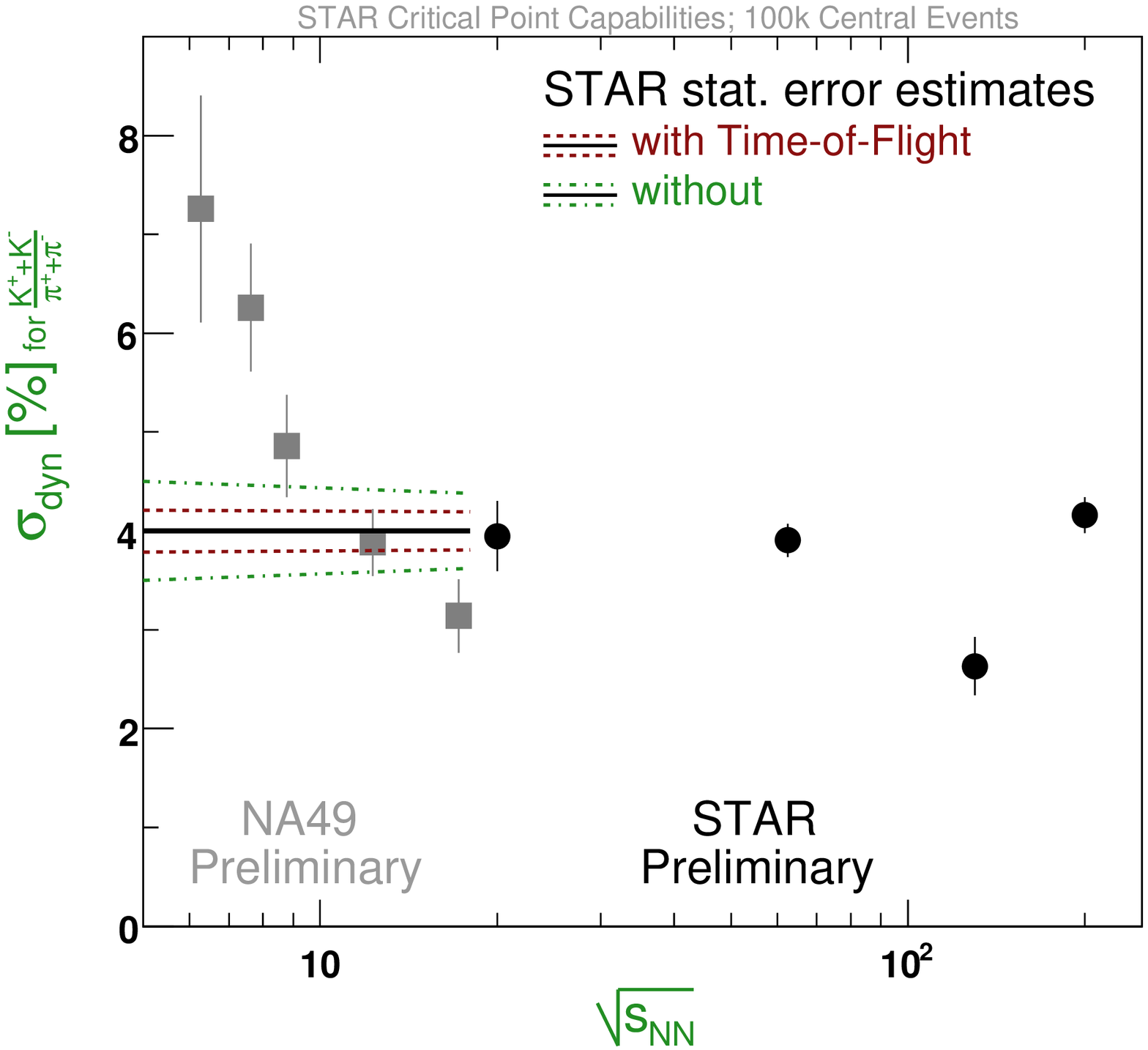}}
\caption{Left panel: Simulation of the event-by-event kaon-to-pion
  ratio for Au+Au collisions at $\sqrt{s_{_{NN}}}=8.76$~GeV with or
  without a Time-of-Flight (TOF) detector available to improve
  particle identification. Right panel: Dynamic fluctuations of the
  kaon-to-pion ratio measured by STAR~\cite{kpianal} and
  NA49~\cite{Roland:2004pu} along with estimates of the errors for
  100,000 central Au+Au collisions with, or without TOF information.}
\label{fig6}
\end{figure}

In Fig.~\ref{fig6} (left) we show simulation results for the
event-by-event distribution of the kaon-to-pion ratio observed by the
STAR detector with or without the TOF upgrade. The simulations are for
100,000 central Au+Au collisions at $\sqrt{s_{_{NN}}}=8.76$~GeV. The
reduction of the size of the kaon sample without the TOF upgrade
causes the distribution to narrow and squeeze against the zero axis:
in many events, no kaons are found. This may make it more difficult to
extract a meaningful width relative to mixed events for the
distribution.

In the right panel of Fig.~\ref{fig6} we show preliminary
STAR~\cite{kpianal} and NA49~\cite{Roland:2004pu} measurements of the
width of the dynamic kaon-to-pion fluctuations: these are extracted
from the difference between the width of real and mixed events. We
also show estimates of the statistical errors expected with the STAR
detector for 100,000 central Au+Au collisions at
$\sqrt{s_{_{NN}}}<18$~GeV. The TOF detector allows the errors to be
reduced by a factor of two compared to STAR without the TOF. The
errors are similar to or smaller than the corresponding NA49
errors. We note however that systematic errors on this measurement are
dominant. In particular, we find that misidentifying 0.5\% of pions as
kaons in our simulations reduces the width of the distribution by
5\%. That systematic uncertainty is as large as the expected
signal. The TOF upgrade is therefore important for this measurement
since it will significantly improve the purity of the kaon sample.

\section{Conclusions}

We've investigated the performance capabilities of the STAR detector
for a low energy scan at RHIC. We find that STAR will be able to
effectively trigger on collisions at these low energies and given the
same number of events will be able to significantly improve on
previous measurements. Measurements taken in a collider geometry will
have smaller $\sqrt{s_{_{NN}}}$-point-to-point systematic errors than
for a fixed target geometry. The efficiency for detecting kaons
however will be smaller due to weak-decays of the kaon.  If TOF
information is not available the available sample of kaons will become
even smaller and dynamical kaon-to-pion fluctuation measurements will
become challenging. Elliptic flow measurements will be significantly
improved because of the STAR detectors large acceptance at
mid-rapidity and symmetric two-$\pi$ azimuthal coverage. Elliptic flow
fluctuation measurements based on the shape of the flow vector
distribution will also be possible. Although not shown here, other
measurements such as $v_1$, HBT, and $\langle p_T \rangle$
fluctuations will be made much better by the STAR detector than by
previous experiments which had smaller acceptances which changed with
$\sqrt{s_{_{NN}}}$.

In summary, a RHIC low energy scan to search for the QCD critical
point will cover a broad region of interest in the nuclear matter
phase diagram. The STAR detector --- a detector designed to measure
the quantities that will be of interest in this search --- will
provide new observables and will improve on previous measurements in
this energy range.


\begin{thebibliography}{99}

\bibitem{whitepapers}
  I.~Arsene {\it et al.}  [BRAHMS Collaboration],
  %``Quark gluon plasma and color glass condensate at RHIC? The perspective
  %from the BRAHMS experiment,''
  Nucl.\ Phys.\ A {\bf 757}, 1 (2005); 
  B.~B.~Back {\it et al.} [PHOBOS Collaboration],
  %``The PHOBOS perspective on discoveries at RHIC,''
  Nucl.\ Phys.\ A {\bf 757}, 28 (2005); 
  J.~Adams {\it et al.}  [STAR Collaboration],
  %``Experimental and theoretical challenges in the search for the quark  gluon
  %plasma: The STAR collaboration's critical assessment of the  evidence from
  %RHIC collisions,''
  Nucl.\ Phys.\ A {\bf 757}, 102 (2005);
  K.~Adcox {\it et al.}  [PHENIX Collaboration],
  %``Formation of dense partonic matter in relativistic nucleus nucleus
  %collisions at RHIC: Experimental evaluation by the PHENIX  collaboration,''
  Nucl.\ Phys.\ A {\bf 757}, 184 (2005). 

\bibitem{cleymans}
  J.~Cleymans, H.~Oeschler, K.~Redlich and S.~Wheaton,
  %``Comparison of chemical freeze-out criteria in heavy-ion collisions,''
  Phys.\ Rev.\ C {\bf 73}, 034905 (2006).

\bibitem{Brown:1990ev}
  F.~R.~Brown {\it et al.},
  %``On the existence of a phase transition for QCD with three light quarks,''
  Phys.\ Rev.\ Lett.\  {\bf 65}, 2491 (1990).

\bibitem{Mocsy:2003qw}
  A.~Mocsy, F.~Sannino and K.~Tuominen,
  %``Confinement versus chiral symmetry,''
  Phys.\ Rev.\ Lett.\  {\bf 92}, 182302 (2004).

\bibitem{Karsch:1998qj}
  F.~Karsch and M.~Lutgemeier,
  %``Deconfinement and chiral symmetry restoration in an SU(3) gauge theory
  %with adjoint fermions,''
  Nucl.\ Phys.\ B {\bf 550}, 449 (1999).

\bibitem{models}
  %\bibitem{Asakawa:1989bq}
  M.~Asakawa and K.~Yazaki,
  %``CHIRAL RESTORATION AT FINITE DENSITY AND TEMPERATURE,''
  Nucl.\ Phys.\ A {\bf 504}, 668 (1989);
  %\bibitem{Barducci:1989wi}
  A.~Barducci, R.~Casalbuoni, S.~De Curtis, R.~Gatto and G.~Pettini,
  %``CHIRAL SYMMETRY BREAKING IN QCD AT FINITE TEMPERATURE AND DENSITY,''
  Phys.\ Lett.\ B {\bf 231}, 463 (1989);
  %\bibitem{Barducci:1989eu}
  A.~Barducci, R.~Casalbuoni, S.~De Curtis, R.~Gatto and G.~Pettini,
  %``CHIRAL PHASE TRANSITIONS IN QCD FOR FINITE TEMPERATURE AND DENSITY,''
  Phys.\ Rev.\ D {\bf 41}, 1610 (1990);
  %\bibitem{Barducci:1993bh}
  A.~Barducci, R.~Casalbuoni, G.~Pettini and R.~Gatto,
  %``Chiral Phases Of QCD At Finite Density And Temperature,''
  Phys.\ Rev.\ D {\bf 49}, 426 (1994);
  %\bibitem{Berges:1998rc}
  J.~Berges and K.~Rajagopal,
  %``Color superconductivity and chiral symmetry restoration at nonzero  baryon
  %density and temperature,''
  Nucl.\ Phys.\ B {\bf 538}, 215 (1999);
  %\bibitem{Halasz:1998qr}
  M.~A.~Halasz, A.~D.~Jackson, R.~E.~Shrock, M.~A.~Stephanov and J.~J.~M.~Verbaarschot,
  %``On the phase diagram of {QCD},''
  Phys.\ Rev.\ D {\bf 58}, 096007 (1998); 
  %\bibitem{Scavenius:2000qd}
  O.~Scavenius, A.~Mocsy, I.~N.~Mishustin and D.~H.~Rischke,
  %``Chiral phase transition within effective models with constituent  quarks,''
  Phys.\ Rev.\ C {\bf 64}, 045202 (2001);
  %\bibitem{Antoniou:2002xq}
  N.~G.~Antoniou and A.~S.~Kapoyannis,
  %``Bootstraping the QCD critical point,''
  Phys.\ Lett.\ B {\bf 563}, 165 (2003);
  %\bibitem{Hatta:2002sj}
  Y.~Hatta and T.~Ikeda,
  %``Universality, the QCD critical / tricritical point and the quark number
  %susceptibility,''
  Phys.\ Rev.\ D {\bf 67}, 014028 (2003).
  
\bibitem{Stephanov:2004wx}
  M.~A.~Stephanov,
  %``QCD phase diagram and the critical point,''
  Prog.\ Theor.\ Phys.\ Suppl.\  {\bf 153}, 139 (2004)
  [Int.\ J.\ Mod.\ Phys.\ A {\bf 20}, 4387 (2005)].

\bibitem{rbrcmeeting} Many details regarding a possible low energy
  scan at RHIC can be found in talks presented at the RIKEN BNL
  Research Center Workshop, ``Can We Discover the QCD Critical Point
  at RHIC'' which can be found at the following web-site:
  https://www.bnl.gov/riken/QCDRhic/talks.asp.

\bibitem{Stephanov:1998dy}
  M.~A.~Stephanov, K.~Rajagopal and E.~V.~Shuryak,
  %``Signatures of the tricritical point in {QCD},''
  Phys.\ Rev.\ Lett.\  {\bf 81}, 4816 (1998)
  [arXiv:hep-ph/9806219].

\bibitem{Ejiri:2005wq}
  S.~Ejiri, F.~Karsch and K.~Redlich,
  %``Hadronic fluctuations at the QCD phase transition,''
  Phys.\ Lett.\ B {\bf 633}, 275 (2006).

\bibitem{Afanasiev:2002mx}
  S.~V.~Afanasiev {\it et al.}  [The NA49 Collaboration],
  %``Energy dependence of pion and kaon production in central Pb + Pb
  %collisions,''
  Phys.\ Rev.\ C {\bf 66}, 054902 (2002). 

\bibitem{Alt:2003ab}
  C.~Alt {\it et al.}  [NA49 Collaboration],
  %``Directed and elliptic flow of charged pions and protons in Pb + Pb
  %collisions at 40-A-GeV and 158-A-GeV,''
  Phys.\ Rev.\ C {\bf 68}, 034903 (2003).

\bibitem{Roland:2004pu}
  C.~Roland {\it et al.}  [NA49 Collaboration],
  %``Event-by-event fluctuations of particle ratios in central Pb + Pb
  %collisions at 20-A-GeV to 158-A-GeV,''
  J.\ Phys.\ G {\bf 30}, S1381 (2004).

\bibitem{Roland:rbrc} G.~Roland ``Experimental Overview and Prospects
  for RHIC,'' RIKEN BNL Research Center Workshop: Can We Discover the
  QCD Critical Point at RHIC?,
  https://www.bnl.gov/riken/QCDRhic/talks.asp.

\bibitem{Ludlam:2003sn}
  T.~Ludlam,
  %``Overview of experiments and detectors at RHIC,''
  Nucl.\ Instrum.\ Meth.\ A {\bf 499}, 428 (2003).

\bibitem{lattTC}
  C.~Bernard {\it et al.}  [MILC Collaboration],
  %``QCD thermodynamics with three flavors of improved staggered quarks,''
  Phys.\ Rev.\ D {\bf 71}, 034504 (2005); 
  M.~Cheng {\it et al.},
  %``The transition temperature in QCD,''
  Phys.\ Rev.\ D {\bf 74}, 054507 (2006); 
  Y.~Aoki, Z.~Fodor, S.~D.~Katz and K.~K.~Szabo,
  %``The QCD transition temperature: Results with physical masses in the
  %continuum limit,''
  Phys.\ Lett.\ B {\bf 643}, 46 (2006). 

\bibitem{lattCP}
  F.~Karsch, C.~R.~Allton, S.~Ejiri, S.~J.~Hands, O.~Kaczmarek, E.~Laermann and C.~Schmidt,
  %``Where is the chiral critical point in 3-flavor QCD?,''
  Nucl.\ Phys.\ Proc.\ Suppl.\  {\bf 129}, 614 (2004);
  Z.~Fodor and S.~D.~Katz,
  %``Critical point of QCD at finite T and mu, lattice results for physical
  %quark masses,''
  JHEP {\bf 0404}, 050 (2004);
  R.~V.~Gavai and S.~Gupta,
  %``The critical end point of QCD,''
  Phys.\ Rev.\ D {\bf 71}, 114014 (2005).

\bibitem{T0} The upper reach of this region is based on estimates of
  the initial temperature achieved in heavy-ion collisions. See e.g.:
   D.~Y.~Peressounko and Yu.~E.~Pokrovsky,
  %``Thermalization temperature in Pb + Pb collisions at SPS energy from  hadron
  %yields and midrapidity p(t) distributions of hadrons and direct
   photons,'' arXiv:hep-ph/0009025;
   U.~W.~Heinz,
  %``'RHIC serves the perfect fluid' - Hydrodynamic flow of the QGP,''
  arXiv:nucl-th/0512051.

\bibitem{STAR}
  K.~H.~Ackermann {\it et al.}  [STAR Collaboration],
  %``STAR detector overview,''
  Nucl.\ Instrum.\ Meth.\ A {\bf 499}, 624 (2003).

\bibitem{tpc}
  M.~Anderson {\it et al.},
  %``The STAR time projection chamber: A unique tool for studying high
  %multiplicity events at RHIC,''
  Nucl.\ Instrum.\ Meth.\ A {\bf 499}, 659 (2003).

\bibitem{UrQMD}
  M.~Bleicher {\it et al.},
  %``Relativistic hadron hadron collisions in the ultra-relativistic quantum
  %molecular dynamics model,''
  J.\ Phys.\ G {\bf 25}, 1859 (1999). 

\bibitem{Shao:2005iu}
  M.~Shao, O.~Y.~Barannikova, X.~Dong, Y.~Fisyak, L.~Ruan, P.~Sorensen and Z.~Xu,
  %``Extensive particle identification with TPC and TOF at the STAR
  %experiment,''
  Nucl.\ Instrum.\ Meth.\ A {\bf 558}, 419 (2006). 

\bibitem{TOF}
  F.~Geurts {\it et al.},
  %``Performance Of The Prototype Mrpc Detector For Star,''
  Nucl.\ Instrum.\ Meth.\ A {\bf 533}, 60 (2004);
  W.~J.~Llope,
  %``The Large-Area Time-Of-Flight Upgrade For Star,''
  Nucl.\ Instrum.\ Meth.\ B {\bf 241}, 306 (2005).

\bibitem{hbt}
  C.~Adler {\it et al.}  [STAR Collaboration],
  %``Pion interferometry of s(NN)**(1/2) = 130-GeV Au + Au collisions at
  %RHIC,''
  Phys.\ Rev.\ Lett.\  {\bf 87}, 082301 (2001).

\bibitem{starv1}
  J.~Adams {\it et al.}  [STAR Collaboration],
  %``Azimuthal anisotropy at RHIC: The first and fourth harmonics,''
  Phys.\ Rev.\ Lett.\  {\bf 92}, 062301 (2004).

\bibitem{balance}
  J.~Adams {\it et al.}  [STAR Collaboration],
  %``Narrowing of the balance function with centrality in Au + Au collisions
  %s(NN)**(1/2) = 130-GeV,''
  Phys.\ Rev.\ Lett.\  {\bf 90}, 172301 (2003). 

\bibitem{starfluct}
  J.~Adams {\it et al.}  [STAR Collaboration],
  %``Multiplicity fluctuations in Au + Au collisions at s(NN)**(1/2) =
  %130-GeV,''
  Phys.\ Rev.\ C {\bf 68}, 044905 (2003);
 J.~Adams {\it et al.}  [STAR Collaboration],
  %``Event-by-event  fluctuations in Au Au collisions at s(NN)**(1/2)  =
  %130-GeV,''
  Phys.\ Rev.\ C {\bf 71}, 064906 (2005);
  J.~Adams {\it et al.}  [STAR Collaboration],
  %``Incident energy dependence of p(t) correlations at RHIC,''
  Phys.\ Rev.\ C {\bf 72}, 044902 (2005);
  J.~Adams {\it et al.}  [STAR Collaboration],
  %``Transverse-momentum p(t) correlations on (eta,Phi) from mean-p(t)
  %fluctuations in Au - Au collisions at s(NN)**(1/2) = 200-GeV,''
  J.\ Phys.\ G {\bf 32}, L37 (2006).

\bibitem{v2meth}
  A.~M.~Poskanzer and S.~A.~Voloshin,
  %``Methods for analyzing anisotropic flow in relativistic nuclear
  %collisions,''
  Phys.\ Rev.\ C {\bf 58}, 1671 (1998). 

\bibitem{starv2}
  K.~H.~Ackermann {\it et al.}  [STAR Collaboration],
  %``Elliptic flow in Au + Au collisions at s(N N)**(1/2) = 130-GeV,''
  Phys.\ Rev.\ Lett.\  {\bf 86}, 402 (2001);
 C.~Adler {\it et al.}  [STAR Collaboration],
  %``Identified particle elliptic flow in Au + Au collisions at  s(NN)**(1/2) =
  %130-GeV,''
  Phys.\ Rev.\ Lett.\  {\bf 87}, 182301 (2001);
  C.~Adler {\it et al.}  [STAR Collaboration],
  %``Azimuthal anisotropy of K0(S) and Lambda + anti-Lambda production at
  %mid-rapidity from Au + Au collisions at s(NN)**(1/2) = 130-GeV,''
  Phys.\ Rev.\ Lett.\  {\bf 89}, 132301 (2002);
  C.~Adler {\it et al.}  [STAR Collaboration],
  %``Elliptic flow from two- and four-particle correlations in Au + Au
  %collisions at s(NN)**(1/2) = 130-GeV,''
  Phys.\ Rev.\ C {\bf 66}, 034904 (2002).

\bibitem{myqm06}
  P.~Sorensen,
  %``Elliptic flow fluctuations in Au + Au collisions at s(NN)**(1/2) =
  %200-GeV,''
  arXiv:nucl-ex/0612021.

\bibitem{qdist}
  J.~Y.~Ollitrault,
  %``On the measurement of azimuthal anisotropies in nucleus--nucleus
  %collisions,''
  arXiv:nucl-ex/9711003;
  N.~Borghini, P.~M.~Dinh and J.~Y.~Ollitrault,
  %``A new method for measuring azimuthal distributions in nucleus nucleus
  %collisions,''
  Phys.\ Rev.\ C {\bf 63}, 054906 (2001).

\bibitem{Alver:2006pn}
  B.~Alver {\it et al.}  [PHOBOS Collaboration],
  %``A method for measuring elliptic flow fluctuations in PHOBOS,''
  arXiv:nucl-ex/0608025.

%\bibitem{Nayak:2006if}
%  T.~K.~Nayak,
  %``Event-by-event fluctuations and the QGP phase transition,''
%  J.\ Phys.\ G {\bf 32}, S187 (2006)

\bibitem{dedx}
  H.~Bichsel,
  %``A method to improve tracking and particle identification in TPCs and
  %silicon detectors,''
  Nucl.\ Instrum.\ Meth.\ A {\bf 562}, 154 (2006).

\bibitem{kpianal}  
  S.~Das  [STAR Collaboration],
  %``Fluctuation studies in STAR,''
  arXiv:nucl-ex/0610044;
  T.~K.~Nayak,
  %``Event-by-event fluctuations and the QGP phase transition,''
  J.\ Phys.\ G {\bf 32}, S187 (2006). 

\end{thebibliography}
\end{document}